\documentclass[amsmath,amssymb,aps,prd,showkeys,reprint, floatfix]{revtex4-1}
\usepackage{graphicx}
\usepackage[breaklinks, colorlinks=true, citecolor=blue]{hyperref}

\newcommand{\GeV}{\,\text{GeV}}
\newcommand{\LQCD}{\Lambda_{\text{QCD}}}
\newcommand{\fr}[2]{\frac{#1}{#2}}

\newcommand{\cN}{\mathcal{N}}
\newcommand{\Tc}{T_c}
\newcommand{\Mkk}{M_{\rm KK}}
\newcommand{\Nf}{N_{\text{f}}}
\newcommand{\Nc}{N_{\text{c}}}
\newcommand{\Ce}{C_e}

\newcommand{\alppr}{\alpha^\prime}

\newcommand{\fux}{f_{ux}}
\newcommand{\fuy}{f_{uy}}
\newcommand{\fuz}{f_{uz}}

\newcommand{\fnx}{f_{0x}}

\newcommand{\cB}{\mathcal{B}}

\newcommand{\rx}{\partial_x}
\newcommand{\ry}{\partial_y}
\newcommand{\rz}{\partial_z}
\newcommand{\rn}{\partial_0}
\newcommand{\ru}{\partial_u}

\newcommand{\rxi}{\partial_\xi}

\newcommand{\bx}{\boldsymbol{x}}

\newcommand{\pr}[1]{\left(#1 \right)}

\begin{document}

\title{Electric conductivity with the magnetic field and the chiral anomaly\\
  in a holographic QCD model}
\author{Kenji Fukushima}
\author{Akitoshi Okutsu}
\affiliation{Department of Physics, The University of Tokyo,
  7-3-1 Hongo, Bunkyo-ku, Tokyo 113-0033, Japan}

\date{\today}

\begin{abstract}
  We calculate the electric conductivity $\sigma$ in deconfined QCD
  matter using a holographic QCD model, i.e., the Sakai-Sugimoto Model
  with varying magnetic field $B$ and chiral anomaly strength.  After
  confirming that our estimated $\sigma$ for $B=0$ is
  consistent with the lattice-QCD results, we study the case with
  $B\neq 0$ in which the coefficient $\alpha$ in the Chern-Simons term
  controls the chiral anomaly strength.  Our results imply that the
  transverse conductivity, $\sigma_\perp$, is suppressed to be
  $\lesssim 70\%$ at $B\sim 1\,\text{GeV}^2$ as compared to the $B=0$
  case when the temperature is fixed as $T= 0.2\,\text{GeV}$.  Since
  the Sakai-Sugimoto Model has massless fermions, the longitudinal
  conductivity, $\sigma_\parallel$, with $B\neq 0$ should diverge due
  to production of the matter chirality.
  Yet, it is possible to extract a regulated part out from
  $\sigma_\parallel$ with an extra condition to neutralize the matter
  chirality.  This regulated quantity is interpreted as an Ohmic part
  of $\sigma_\parallel$.  We show that the longitudinal Ohmic
  conductivity increases with increasing $B$ for small $\alpha$, while
  it is suppressed with larger $B$ for physical $\alpha=3/4$ due
  to anomaly induced interactions.
\end{abstract}

\maketitle

\section{Introduction}


Chiral anomaly has been a profitable probe to the nonperturbative
sector of quantum field theories such as quantum chromodynamics (QCD)
since its discovery dated back in the
'60s~\cite{Bell:1969ts, *Adler:1969gk}.  The resolution of the PCAC
(partially conserved axial current) puzzle~\cite{Sutherland:1967vf}
via anomalous $\pi^0\to \gamma\gamma$ is the most well known example.
The virtue of the chiral anomaly is not limited to a specific
calculation of the decay rate, and the chiral anomaly has gone on to
manifest itself in various QCD phenomena.  Another famous example is
the $\eta$-$\eta^\prime$ puzzle or the $\mathrm{U(1)_A}$ puzzle; that
is, the $\eta^\prime$ mass is significantly heavier than other pseudo
scalar mesons belonging to the same nonet.  It was 't~Hooft who gave
the explanation by the instanton mechanism~\cite{tHooft:1976rip}
associated with the chiral anomaly.  A more recent example is found in
discussions on the QCD phase diagram and a hypothetical critical
point may emerge from the quark-quark and quark-antiquark coupling
induced by the chiral anomaly~\cite{Hatsuda:2006ps, *Abuki:2010jq}.
We should emphasize that the chiral anomaly exists not only in QCD but
in a wider class of gauge theories with chiral fermions.  The
establishment of three dimensional materials with relativistic
fermionic dispersions, namely, the Weyl semimetals and the Dirac
semimetals, has expanded the relevance of the chiral anomaly to
physics of condensed matter.


In QCD it has been a longstanding problem how to reveal topologically
nontrivial aspects of the QCD vacuum experimentally, though we are
theoretically familiar with the QCD vacuum structure well;  for
example, Dashen's phenomenon~\cite{Dashen:1970et} is textbook
knowledge but there is no way to verify it by nuclear experiments.
Along these lines the Chiral Magnetic Effect (CME) has been proposed
as an experimental signature of the chiral anomaly detected in nuclear
experiments~\cite{Kharzeev:2007jp}, which was formulated in terms of
the chiral chemical potential $\mu_5$ field-theoretically
later~\cite{Fukushima:2008xe}.  The CME predicts anomalous generation
of the electric current in parallel to an applied magnetic field $B$
(see Ref.~\cite{Gursoy:2014aka} for magnetic effects in the heavy-ion
collision experiments) if a medium has imbalanced chirality;  see also
Refs.~\cite{Miransky:2015ava, Fukushima:2018grm} for reviews on
nontrivial magnetic field effects.
Interestingly, the electric current is anomaly protected and not
renormalized with interactions, which also indicates that the current is
nondissipative.  We can understand this nondissipative nature from the
fact that the coefficient in response to $B$ is $T$ even, making a
sharp contrast to the Ohmic conductivity which is $T$ odd since the
electric field $E$ has the time reversal property opposite to $B$.
Interested readers can consult a comprehensive
review~\cite{Kharzeev:2015znc} for theoretical backgrounds and
experimental prospects of the CME and related chiral effects such as
the Chiral Separation Effect, the Chiral Vortical Effect, etc (see also
Ref.~\cite{Kharzeev:2020jxw} for the ongoing projects of the nuclear
experiments).


It was an ingenious idea that the electric conductivity could exhibit
characteristic dependence on the magnetic field to signify the CME:
the longitudinal electric conductivity, $\sigma_\parallel$, involves
chirality production due to parallel $E$ and $B$ and the produced
chirality gives rise to the CME in response to $B$ again.  Thus, in
total, $\sigma_\parallel$ is expected to have a CME induced
contribution that increases as $\propto B^2$.  This increasing
behavior of $\sigma_\parallel\propto B^2$ (i.e., the decreasing
behavior of the resistance $\sim \sigma_\parallel^{-1}\propto B^{-2}$) is referred
to as the negative magnetoresistance~\cite{Son:2012bg}.  In fact, in
condensed matter systems, the negative magnetoresistance has been
observed and it is believed that the CME has an experimental
confirmation, as first reported in Ref.~\cite{Li:2014bha}, but there
are two major gaps between theory and experiment.


First, for undoubted establishment, it is crucial to make reliable
theory estimates for the anomaly related $\sigma_\parallel$.  In both
Refs.~\cite{Son:2012bg,Li:2014bha} the relaxation time approximation
was employed with a strong assumption that the relaxation time is $B$
independent.  In principle fermion interactions at the microscopic
level may have significant $B$ dependence which may in turn change the
$B$ dependence of $\sigma_\parallel$.  Theoretical studies hitherto
only concentrated on extreme regions of the parameters.  In the strong
magnetic field regime at high enough temperature $T$, perturbative QCD
calculations become feasible in the lowest Landau level (LLL)
approximation~\cite{Hattori:2016cnt, *Hattori:2016lqx}.  This LLL
approximation was relaxed later and the full Landau level sum was
taken in Ref.~\cite{Fukushima:2017lvb, *Fukushima:2019ugr}.  However,
high $T$ is still required to justify the weak coupling treatment.
In QCD the asymptotic freedom guarantees the validity of the weak
coupling treatment at high $T$, but the condensed matter system may
not have such a property.   We therefore need to gain some insights
into nonperturbative computations.


Second, it would be indispensable to quantify the $B$ dependence in
other pieces of the electric conductivity.  Because the fermion
interactions are generally affected by $B$, even without the chiral
anomaly, the  Ohmic part of $\sigma_\parallel$ should be $B$
dependent.  If its dependence looks similar to what is expected from
the chiral anomaly, the physical interpretation of the negative
magnetoresistance would become subtle.  We should then discuss not
only whether the $B$ dependence is positive or negative, but we must
go through quantitative comparisons.  Besides, it is anyway an
interesting theory question how the Ohmic $\sigma_\parallel$ may be
changed by the chiral anomaly.  If the chirality production process is
discarded (which will be discussed in this work), no CME contribution
of $\sigma_\parallel \propto B^2$ arises, but still a finite Ohmic part
can  be sensitive to the strength of the chiral anomaly.  One might
think that the anomaly is dictated by the theory and it is unrealistic
to control its strength.  In the holographic QCD model, as we will
explain soon later, we have a parameter to control it.  Intuitively,
this manipulation of changing
the anomaly parameter is analogous to exploring the effect of the
$\mathrm{U(1)_A}$ breaking interaction in QCD, i.e., the
Kobayashi-Maskawa-'t~Hooft interaction in chiral effective models.
For example, the fate of the QCD Critical Point may differ depending
on how much the effective restoration of $\mathrm{U(1)_A}$ symmetry
occurs~\cite{Fukushima:2008wg, *Chen:2009gv}.  The point is that the
axial Ward identity itself is intact, but its expectation value could be
changed by the instanton density accommodated by the considered
state~\cite{Shuryak:1993ee}.  In principle it should arise from
systematic calculations, but we will use a probe approximation in
which back reactions are dropped.  In this approximation we do not
know how much the chirality production would be affected by medium
effects, and we will emulate such effects by modifying a control
parameter.


For our present purpose to resolve the aforementioned problems, the
Sakai-Sugimoto Model (SSM) is an advantageous
choice~\cite{Sakai:2004cn} (see Ref.~\cite{Rebhan:2014rxa} for a
review for nuclear physicists, and Ref.~\cite{Gursoy:2021efc} for
a review on holographic QCD matter at finite $B$, and also
Ref.~\cite{Nakas:2020hyo} for a related work on the baryon operator).
The SSM is one of the most established holographic QCD models and it
has the same low-energy
degrees of freedom as QCD\@.  The most notable feature of the SSM is
that it realizes exactly the same pattern of the chiral symmetry
breaking. This model has been quite successful in reproducing the
hadron spectrum as shown in the original proposal~\cite{Sakai:2004cn},
and the applications cover even the glueball
physics~\cite{Brunner:2015oqa, *Brunner:2015yha}.  In the context of
the CME physics, the chiral magnetic conductivity in the presence of
$\mu_5$ was first reproduced correctly in Ref.~\cite{Yee:2009vw}, but
the treatment of $\mu_5$ caused some problematic complications, which
was clearly pointed out in Ref.~\cite{Rebhan:2009vc}.  In contrast the
negative magnetoresistance would not need $\mu_5$, and what should be
calculated is the $B$ dependence of $\sigma_\parallel$.
Interestingly, there are preceding works, Ref.~\cite{Bergman:2008sg}
at $B=0$ and Ref.~\cite{Lifschytz:2009si} at $B\neq 0$, for the
holographic conductivity estimates in the SSM\@.  See
Ref.~\cite{Li:2018ufq, Bu:2018psl, *Bu:2019mow} for discussions in a different
holographic setup.  We also mention that nontrivial behavior of the
magnetoresistance (including a positive value) has been found by
holographic calculations in
Refs.~\cite{Landsteiner:2014vua, *Jimenez-Alba:2015awa, Sun:2016gpy}.
A more recent calculation of the conductivity in a
Einstein-dilaton-three-Maxwell holographic model is found in
Ref.~\cite{Arefeva:2021jpa}.
It is highly nontrivial to compare our results to
preceding ones due to model and convention differences, but a solid
conclusion of Ref.~\cite{Lifschytz:2009si} is that the longitudinal
conductivity diverges with massless fermions.  This is naturally
understood within the framework of the SSM;  in the massless theory
the produced chirality does not decay.  So, the chirality charge
increases proportional to the time $t$, and the CME current increases
also as $\propto t$, which makes
the conductivity at the zero frequency limit diverge.  Thus, the
divergence of the conductivity is a physically sensible behavior.
It should be noted that the chirality can in principle relax even in
the massless theory~\cite{Iatrakis:2015fma}, but its relaxation time
scale scales as $\Nc$ and in our calculation at the leading-order in
the large-$\Nc$ limit this effect is negligible.
In the present work we argue that a particular choice of an additional
condition allows us to extract a finite piece of the conductivity,
which, we interpret, is an Ohmic part.  Here, our considerations are
limited to QCD matter, but we mention that there are already proposals
for holographic models for the Weyl semimetals~\cite{Gursoy:2012ie,
  Jacobs:2014nia, Rogatko:2017svr,  Landsteiner:2019kxb} and our
methodology should be applicable there for further investigations.

This paper is organized as follows.  In Sec.~\ref{sec:formulation} we
make a brief overview of the SSM introducing some notations and
explain how the chiral anomaly is implemented in the model.  We will
convert all the expressions in the physical units in the end, and we
will here make clear model parameters.  In Sec.~\ref{sec:zero} we will
present the $B=0$ result and compare it to the lattice-QCD
calculations.  In Sec.~\ref{sec:finite} we will generalize our
calculations to the finite-$B$ case and we will see that the
transverse conductivity is suppressed by large $B$ as physically
expected.  Our main finding is that the finite part of the
longitudinal conductivity has nonmonotonic and complicated dependence
on $B$ and the anomaly parameter.  We will summarize our discussions
in Sec.~\ref{sec:summary}.

\section{Formulation}
\label{sec:formulation}

The Sakai-Sugimoto model consists of $\Nf$ D8/$\overline{\rm D8}$
branes for the left- and the right-handed quarks, respectively, and
$\Nc$ D4 branes for the gluons wrapped around an $S^1$ of radius
$\Mkk^{-1}$ in the $x_4$ direction~\cite{Sakai:2004cn}.  Because of
the presence of $\Mkk$ which eventually corresponds to $\LQCD$ in QCD,
conformality is lost, and a periodic boundary condition for bosons and
antiperiodic condition for fermions in the $x_4$ direction break
supersymmetry.  The massless quarks exhibit chiral symmetry identical
to QCD symmetry, i.e.,
$\mathrm{U}(N_f)_{\rm L}\times \mathrm{U}(N_f)_{\rm R}$ among which
$\mathrm{U(1)_A}$ is broken by the axial anomaly.  When the two D8
branes are parallelly separated, the model represents chiral symmetry
restoration in the deconfined phase, while the confining geometry
inevitably leads to chiral symmetry breaking in this model and this
feature is consistent with the expected interplay between chiral
symmetry and confinement.  In the present work we will consider only
the chiral symmetric case at $T>\Tc$ where $\Tc=\Mkk/(2\pi)$
represents a critical temperature of the confinement-deconfinement
phase transition.

In the SSM on the D8 brane an effective five-dimensional form with the
Dirac-Born-Infeld (DBI) action and the Chern-Simons term accounting
for the topological
source~\cite{Sakai:2004cn, Yee:2009vw, Rebhan:2014rxa}
are considered as
\begin{align}
  S
  &= S^{\mathrm{YM}} + S^{\mathrm{CS}}  \notag \\
    &= \cN   \int d^4x\, du \biggl[- u^{\fr{1}{4}}
      \sqrt{-\mathrm{det}\pr{g_{\alpha\beta}+F_{\alpha\beta}}} \notag\\
    &\qquad\qquad\qquad\qquad
          + \frac{\alpha}{3!} \epsilon^{\mu\nu\rho\sigma\tau}
    A_{\mu} F_{\nu\rho} F_{\sigma\tau} \biggr]\,.
    \label{eq:action}
\end{align}
We note that all the coordinates and the gauge fields are rescaled by
the AdS radius $R$ as
\begin{equation}
  x = \fr{\tilde{x}}{R}\,,\qquad
  A = \fr{2\pi\alppr}{R} \tilde{A}\,,\qquad
  F = 2\pi\alppr \tilde{F}\,,
  \label{eq:rescale}
\end{equation}
where tilde quantities are original variables with mass dimensions and
$\alpha'=l_s^2$ with $l_s$ being the string length scale.  The overall
normalization constant is given by
\begin{equation}
  \cN = \frac{\Nf \Nc R^6}{12\pi^2 (2\pi \alpha')^3}\,.
\end{equation}
It would be convenient to express a combination of $\alpha'$ and $R$
in terms of the Kaluza-Klein mass through
\begin{equation}
  \Mkk = \frac{\lambda \alpha'}{2 R^3}\,,
\end{equation}
where $\lambda=g^2 \Nc$ is the 't~Hooft coupling.  We also note that
the Chern-Simons coefficient $\alpha$ in Eq.~\eqref{eq:action} is
fixed as $\alpha=3/4$, which stems from a compact expression of
$S^{\mathrm{CS}}=(N_c/24\pi^2)\int \omega_5(\tilde{A})$.  In our
expression the $u$-integration in Eq.~\eqref{eq:action} runs only on
D8, and the overall coefficient should be doubled including the
$\overline{D8}$ contribution.

For $T>\Tc$ the induced metric on the D8 brane takes the following
form:
\begin{equation}
  ds^2 = u^{3/2}\bigl[ f(u) d\tau^2 + d\bx^2\bigr]
  + \Bigl[u^{3/2} x_4'(u)^2 + \frac{1}{u^{3/2}f(u)} \Bigr] du^2\,,
\end{equation}
where $u$ denotes the radial coordinate transverse to the D4 branes
and $f(u)=1-(u_T/u)^3$ with $u_T=(4\pi/3)^2 R^2 T^2$.  For our purpose
to estimate the electric conductivity in the linear response regime,
we need to keep the (time-dependent) vector potential $a_{x,y,z}$ up
to the quadratic order in the action, while we should retain full $B$
dependence to cover the scope of strong $B$ regions.  Without loss of
generality we can choose $B$ along the $z$ axis.  Then, in the $a_u=0$
gauge, the explicit form of rescaled
$g_{\alpha\beta} + F_{\alpha\beta}$ reads:
\begin{equation}
  \begin{split}
  & g_{\alpha\beta} + F_{\alpha\beta} \\
  & \simeq  \begin{pmatrix}
-u^{3/2}f(u) & f_{0x} & f_{0y}  & f_{0z} & -a_0' \\
-f_{0x} & u^{3/2} & B  & 0 & -a_x' \\
-f_{0y} & -B & u^{3/2} &  0 & -a_y' \\
-f_{0z} & 0  & 0  & u^{3/2} & -a_z' \\
a_0' & a_x' & a_y' & a_z' & \displaystyle \frac{1}{u^{3/2}f(u)}
\end{pmatrix}\,.
\end{split}
\end{equation}
We note that the prime represents $\partial/\partial u$.  We should
solve the equations of motion (EoMs) for $a_{x,y,z}$ with boundary
condition that we will explain later.  For $a_x$ the EoM is, with the
action~\eqref{eq:action}, given as
\begin{equation}
  \ry \frac{\delta S}{\delta B}
  -\rn \frac{\delta S}{\delta \fnx}
  -\ru \frac{\delta S}{\delta a_x'} = 0\,.
\label{eq:eom_ax}
\end{equation}
For this case of $a_x$ the Chern-Simons action, $S^{\mathrm{CS}}$,
produces only higher order terms beyond the linear response regime.
Since we consider a spatially homogeneous situation only, we can
safely drop the spatial derivative and simplify the differential
equations.  Below we will drop terms involving $\rx$, $\ry$, and
$\rz$.  The EoM for $a_y$ has a similar structure with $x$ replaced
with $y$.  Because of the presence of $B$, the EoM for $a_z$ has an
extra contribution from the Chern-Simons action as
\begin{equation}
  -\rn \frac{\delta S}{\delta f_{0z}}
  -\ru \frac{\delta S}{\delta a_z'}
  + \frac{\delta S}{\delta a_z} = 0\,.
\end{equation}
The last term arises from $S^{\mathrm{CS}}$ which is proportional to
$F_{xy}F_{u0}\sim B a_0'$.  These EoMs should be coupled with a
constraint from $a_0$, i.e.,
\begin{equation}
  - \ru \frac{\delta S}{\delta a_0'}
  + \frac{\delta S}{\delta a_0} = 0\,.
\label{eq:eom_a0}
\end{equation}
Again, the Chern-Simons action yields the last term which is proportional
to $F_{xy}F_{zu}\sim B a_z'$.  Also, one more constraint appears from
$a_u$ (which is needed even for the $a_u=0$ gauge in a way analogous
to the Gauss law even in the Weyl gauge), that is,
\begin{equation}
   -\rn \frac{\delta S}{\delta \partial_0  a_u}
  + \frac{\delta S}{\delta a_u} = 0\,.
\label{eq:eom_au}
\end{equation}
Here, because the action depends on $a_u$ only through
$f_{0u}=\partial_0 a_u-\partial_u a_0$, we can replace
$\delta S/\delta \partial_0 a_u$ with $-\delta S/\delta a_0'$.
This last constraint is quite interesting from the point of view of
the chiral anomaly.  In fact, we can identify the current from
\begin{equation}
  j_{L/R}^\mu = \mp \frac{\delta S^{\mathrm{YM}}}
  {\delta \partial_u \tilde{A}_\mu} \Bigr|_{u=\pm\infty}\,.
\end{equation}
Here, our notation may look a little sloppy;  in the
action~\eqref{eq:action} $u$ runs only on D8, but in the above concise
expression $u$ is extended toward $-\infty$.  It is easy to find,
\begin{equation}
  \frac{\delta S^{\mathrm{CS}}}{\delta \tilde{A}_u} =
  \frac{\Nc\Nf}{32\pi^2} \epsilon^{\mu\nu\rho\sigma}
  \tilde{F}_{\mu\nu} \tilde{F}_{\rho\sigma}\,,
\end{equation}
where we took account of the derivatives of $\tilde{A}_\mu$ via the
integration by part [and there is no need to consider contributions
from the first term in Eq.~\eqref{eq:eom_au}].  The flavor $\Nf$
appears from the trace in $\omega_5(\tilde{A})$.  Therefore, adding
both left and right sectors up, Eq.~\eqref{eq:eom_au} at
$|u|\to\infty$ immediately recovers (dropping all spatial
derivatives):
\begin{equation}
  \partial_0 n_5 = -\frac{\Nc\Nf}{16\pi^2}\epsilon^{\mu\nu\rho\sigma}
  \tilde{F}_{\mu\nu}\tilde{F}_{\rho\sigma}
\end{equation}
in the case without axial vector components.  For the setup with axial
vector fields, in contrast, more careful treatments are crucial as
discussed in Refs.~\cite{Rebhan:2008ur,Rebhan:2009vc}.  For the
present purpose within only the vector gauge fields, this simple
identification of Eq.~\eqref{eq:eom_au} as the chiral anomaly works
straightforwardly.

For later convenience, though it is a little lengthy expression, we
shall write down the expanded form of $S^{\mathrm{YM}}$ up to the
quadratic order, i.e.,
\begin{align}
  & \sqrt{-\det(g_{\alpha\beta} + F_{\alpha\beta})}
  \simeq u^{9/4}\sqrt{\cB}
  + \frac{u^{9/4}}{2\sqrt{\cB}} \bigl[ f(a_x^{\prime 2} +a_y^{\prime 2}) \notag\\
  &\quad  - \cB (a_0^{\prime 2} - f a_z^{\prime 2}) \bigr]
   - \frac{u^{-3/4} (f_{0x}^2+f_{0y}^2 + \cB f_{0z}^2)}{2 f \sqrt{\cB}}\,,
 \end{align}
where we introduced a short-hand notation (see also
Ref.~\cite{Fukushima:2013zga} for notation) as
\begin{equation}
 \cB = 1+B^2 u^{-3} \,.
\end{equation}
With these expressions and notations, we are ready to proceed to
concrete calculations of the electric conductivity.


\section{Zero magnetic field case and consistency check with the
  lattice-QCD estimate}
\label{sec:zero}

Before considering the full magnetic dependence, let us solve these
equations in a much simpler case at $B=0$.  This exercise would be
useful to explain the procedures in a plain manner, and also, we can
make a quantitative comparison to the electric conductivity from the
lattice-QCD results which are available for the $B=0$ case only.

In this case of $B=0$ all the contributions from the Chern-Simons
terms are simply dropped off and also the DBI action significantly
simplifies with $\cB=1$.  Then, the constraints~\eqref{eq:eom_a0} and
\eqref{eq:eom_au} become,
\begin{equation}
  -\ru \bigl( u^{5/2} a_0' \bigr) = 0\,,\qquad
  \rn \bigl( u^{5/2} a_0' \bigr) = 0\,.
  \end{equation}
One obvious solution is $a_0' = c\,u^{-5/2}$ with a constant $c$.
We recall that our calculations are at finite $T$ but zero chemical
potential, so the density should be zero leading to $c=0$ and then
$a_0=0$ is entirely chosen.

In the absence of $B$, there is no preferred direction and all EoMs
for $a_{x,y,z}$ are equivalent.  A simple calculation gives,
\begin{equation}
  -u^{-1/2} f^{-1} \rn f_{0i} + \ru \bigl( u^{5/2} f a_i' \bigr) = 0\,.
\end{equation}
Under coordinate transformation $\xi = u_T \mathrm{/} u$ and using the
Fourier transformed variable $a_i(\xi,\omega)$, we can rewrite the EoM as
\begin{equation}
  \xi^{-3/2} \frac{\Omega^2}{1-\xi^3} a_i
  + \rxi \Bigl[ \xi^{-1/2}
  (1-\xi^3)
  \rxi a_i \Bigr] = 0\,.
  \label{eq:eom_zerob}
\end{equation}
Here, we defined the dimensionless frequency as
$\Omega^2 = \omega^2\mathrm{/}u_T$.

As discussed in Ref.~\cite{Son:2002sd}, we should impose the infalling
boundary condition near the blackhole horizon at $u\sim u_T$ or
$\xi\sim 1$.  We can approximate the EoM near $\xi\sim 1$ and
identify the asymptotic form of the solution from
\begin{equation}
  \frac{\Omega^2}{3(1-\xi)} a_i - 3 \rxi a_i + 3(1-\xi) \rxi^2 a_i =  0\,,
\label{eq:eom_zerob_asym}
\end{equation}
which is obtained from Eq.~\eqref{eq:eom_zerob} near $\xi\sim 1$.  We
can easily solve Eq.~\eqref{eq:eom_zerob_asym} using the asymptotic
form, $a_i \sim (1-\xi)^\delta$, from which
$(\Omega/3)^2+\delta+\delta(\delta-1)=0$ follows, leading to
$\delta=\pm \frac{i\Omega}{3}$ immediately.  The infalling direction
corresponds to $\delta=-\frac{i\Omega}{3}$ and we can parametrize the
solution as
\begin{align}
  a_i(\xi) = \pr{1-\xi}^{-\fr{i\Omega}{3}} g(\xi)\,,
\end{align}
where $g(\xi)$ is a regular function near $\xi\sim 1$.   The
normalization of $a_i(\xi)$ is conventionally chosen as the unity, i.e.,
$a_i(\xi=0)=1$ or $g(\xi=0)=1$.  We can then expand $g(\xi)$ for
small $\Omega$, under the condition that $g(\xi=1)$ is regular.  Up to
the first order in $\Omega$ we can drop the first term in
Eq.~\eqref{eq:eom_zerob} and the equation to be satisfied by $a_i$ is
\begin{equation}
  \rxi \Bigl[ \xi^{-1/2}
  (1-\xi^3)
  \rxi a_i \Bigr] = 0\,,
  \label{eq:reduced_eq}
\end{equation}
which can be solved as
\begin{equation}
  a_i(\xi) = C \int_0^\xi d\xi\,\frac{\xi^{1/2}}{1-\xi^3} + D
  = \frac{C}{3}\ln\biggl(\frac{1+\xi^{3/2}}{1-\xi^{3/2}}\biggr) + D\,,
\end{equation}
where $C$ and $D$ are $\Omega$ dependent constants.  We can then write
down a form of $g(\xi)$ for small $\Omega$ as
$g(\xi)\simeq [1+i\frac{\Omega}{3}\ln(1-\xi)] a_i(\xi)$.  The
condition of $a_i(\xi=0)=1$ fixes $D=1$, and the regularity of
$g(\xi\to 1)$ fixes $C = i\Omega$.  Therefore, we can conclude,
\begin{equation}
  g(\xi) = 1 + \frac{i\Omega}{3} \ln\biggl[
  \frac{(1-\xi)(1+\xi^{3/2})}{1-\xi^{3/2}} \biggr] + O(\Omega^2)\,.
\end{equation}
In response to the boundary condition at the infrared (IR) side, the
behavior at the ultraviolet (UV) side near $\xi\sim 0$ is fixed, from
which the physical information can be extracted.  That is,
\begin{equation}
  a_i(\xi) \simeq 1 + \frac{2i\Omega}{3}\xi^{3/2} +\cdots
  \label{eq:a_expand}
\end{equation}
Now, let us prescribe how to calculate the electric current
expectation value using the GKP-W relation
\cite{Gubser:1998bc, Witten:1998qj}.  It is the generating functional
coupled to the gauge potential, which results from the on-shell action
in the gravity theory with the UV boundary condition of $a_i(\xi\to
0)$ as the physical vector potential in the gauge theory.

To calculate the electric current expectation value, thus, we should
take a functional derivative of the gravity action with respect to
$a_i$ on the UV boundary.  Near the UV boundary ($\xi\sim 0$ or
$u\sim \infty$), the action has asymptotic behavior as follows:
\begin{align}
  S &\sim -\cN \int d^4 x du \, u^{5/2}\, \fr{1}{2}
      (\fux^2 + \fuy^2 + \fuz^2) \notag\\
    &\sim  -\fr{\cN u_T^{3/2}}{2} \int d^4 x d\xi \,
      \xi^{-\fr{1}{2}} (\rxi a_i)^2 \,.
\end{align}
Therefore, the dimensionless electric current is
\begin{align}
  j_i &= \fr{\delta S}{\delta \partial_\xi a_i(\xi=0)} \notag\\ 
  &= -2 \biggl(-\fr{\cN u_T^{\fr{3}{2}}}{2} \biggr)\,
  \xi^{-\fr{1}{2}} \, \rxi a_i \Bigr|_{\xi=0}
  = i\cN\omega u_T \,.
\end{align}
This is an expression in the dimensionless units.  We note that
$j_i=\sigma E_i$ translates to $j_i=i \sigma \omega A_i$ in frequency
space (if $\sigma$ is a time-independent constant).  We note that our
normalization is $a_i(\xi\to 0)=1$ and we should add the
$\overline{D8}$ contribution multiplying a factor $2$.  Plugging
$u_T=(4\pi/3)^2 R^2 T^2$ into $j_i$, we can derive the electric
conductivity:
\begin{equation}
  \frac{\sigma}{q^2} = 2\biggl(\fr{4\pi}{3}\biggr)^2 \cN (2\pi\alppr)^2 R^{-3}T^2
  = \frac{2\lambda \Nf \Nc T^2}{27\pi \Mkk} \,.
\end{equation}
Here, we retrieved $2\pi\alpha'$ from Eq.~\eqref{eq:rescale} and also
recovered the electric charge $q$.  This $T^2$ behavior is consistent
with preceding studies, see Ref.~\cite{Bergman:2008sg}.

Once the t'~Hooft coupling, $\lambda$, and the Kaluza-Klein mass,
$\Mkk$, are determined to reproduce the physical quantities, we
can express $\sigma$ in the physical units.  More specifically, the
$\rho$ meson mass, $m_\rho$, and the pion decay constant, $f_\pi$, can
fix these parameters as~\cite{Rebhan:2014rxa,Sakai:2005yt,Sakai:2004cn}
\begin{equation}
  \lambda = 16.63\,,\qquad M_{KK} = 0.95\GeV \,.
\end{equation}

\begin{table}
\centering
\begin{tabular}{|c||c|c|c|}
\hline
 $\sigma/(\Ce T)$  & 1.1$\Tc$  & 1.3$\Tc$  & 1.5$\Tc$  \\ \hline
This work & 0.206       & 0.243       & 0.281       \\ \hline
Lattice-QCD~\cite{Ding:2016hua}    & ~0.201-0.703~ & ~0.203-0.388~ & ~0.218-0.413~ \\ \hline
\end{tabular}
\caption{Comparison between our estimates and the lattice-QCD results
  from Ref.~\cite{Ding:2016hua} for the dimensionless electric
  conductivity for three different temperatures above $\Tc$.}
\label{tab:comparison}
\end{table}

To make a quantitative comparison to the lattice-QCD results for
$\Nf=2$, we should consider normalized $\sigma$ by the flavor factor,
$\Ce=(2e/3)^2+(-e/3)^2=5e^2/9$.  In our calculation we simply treated
the electric charge in the normalization, which implies that the above
expression is already normalized.  Then,
\begin{equation}
  \frac{\sigma}{\Ce T} = \frac{2\lambda \Nc T}{27\pi \Mkk}
  = \frac{\lambda}{9\pi^2} \biggl(\frac{T}{\Tc}\biggr)\,,
\label{eq:final_zerob}
\end{equation}
where we used the known relation $\Tc=\Mkk/(2\pi)$ in the SSM\@.
Table~\ref{tab:comparison} shows the comparison between our SSM
estimates and the lattice-QCD results from Ref.~\cite{Ding:2016hua},
indicating good agreement.  Our estimates also match with the more 
recent results in Ref.~\cite{Astrakhantsev:2019zkr}.
We should, however, not take the comparison too seriously.
The probe approximation is justified only
in the $\Nc\to\infty$ limit, but QCD has only $\Nc=3$, and $1/\Nc$
corrections are expected.

\section{Finite magnetic case}
\label{sec:finite}

We can repeat the same procedures including full $B$ effects.  First,
let us consider the transverse degrees of freedom, i.e., the $x$ and
$y$ directions perpendicular to $B$.  For $a_{x,y}$, as seen from
Eq.~\eqref{eq:eom_ax}, there is no contribution from the Chern-Simons
term and the analysis is easier than the longitudinal direction.  The
EoM for $a_x$ is,
\begin{equation}
  -u^{-1/2} f^{-1} \cB^{-1/2} \rn f_{0x}
  +\ru \bigl( u^{5/2} f \cB^{-1/2} a_x' \bigr) = 0\,.
\end{equation}
In the same way as the $B=0$ case, in frequency space and in
terms of $\xi=u_T/u$, we can rewrite the above into
\begin{equation}
  \xi^{-3/2} \frac{\Omega^2}{1-\xi^3} \cB^{-1/2} a_x
  + \rxi \bigl[ \xi^{-1/2} (1-\xi^3)
  \cB^{-1/2} \rxi a_x \bigr] = 0\,.
\end{equation}
Near $\xi\sim 1$, the asymptotic behavior is determined by the
singular part of the above EoM which is the same as the $B=0$ case.
Then, we can take the form of the solution to be
$a_x(\xi)=(1-\xi)^{-\frac{i\Omega}{3}}g(\xi)$ and expand $g(\xi)$ for
small $\Omega$.  Some calculations similar to previous ones lead us to
the following solution:
\begin{equation}
  a_x(\xi) = C\int_0^\xi d\xi\,\frac{\xi^{1/2} \cB^{1/2}}{1-\xi^3} + D\,.
\label{eq:ax_integ}
\end{equation}
The integration is analytically possible but the expression is highly
intricate.  Nevertheless, the previous exercise at $B=0$ tells us that
$C$ is fixed to cancel the singularity of $\ln(1-\xi)$ around
$\xi\sim 1$, which requires,
\begin{equation}
  C = i\Omega\,\cB_0^{-1/2}\,,\qquad
  D = 1\,,
\end{equation}
where $\cB_0=1+B^2 u_T^{-3}$.  Once these constants are known, we can
expand $a_x(\xi)$ near $\xi\sim 0$ as
\begin{equation}
  a_x(\xi) \simeq 1+\frac{2i\Omega}{3} \cB_0^{-1/2} \xi^{3/2}+\cdots
\end{equation}
Therefore, the correction due to $B$ is simply $\cB_0^{-1/2}$ and the
conductivity is, thus,
\begin{equation}
  \sigma_\perp = \frac{\sigma(B=0)}{\sqrt{1+B^2 u_T^{-3}}}\,,
  \label{eq:s_mag_t}
\end{equation}
where $\sigma(B=0)$ is given by Eq.~\eqref{eq:final_zerob}.  The
transverse conductivity is suppressed by large $B$, and this makes
physical sense.  The external magnetic field restricts the transverse
motion of charged particles and the charge transport along the
transverse directions needs a jump between different Landau levels.
In the strong $B$ limit, therefore, the electric conductivity should
be vanishing.
We note that the drift motion of charged particles under $B$ may
change the scenario.  In the probe appxomation of the SSM the drift
motion effect (whose time scale $\propto\Nc/\Nf$) is negligible and
our calculations are justified.  For the AC 
conductivity, for which the drift frequency can be smaller than the
electric frequency, the transverse conductivity should not be
vanishing even in the strong $B$ limit, see Ref.~\cite{Li:2018ufq} for
details.

Next, we shall find the longitudinal conductivity.  To this end we
consider the constraints and then solve the EoMs as we did for the
$B=0$ case.  From Eq.~\eqref{eq:eom_a0} we have
\begin{equation}
  -\ru (u^{5/2} \cB^{1/2} a_0') - 4\alpha B a_z' = 0\,,
\end{equation}
which means that $u^{5/2} \cB^{1/2} a_0'+4\alpha B a_z$ is a $u$
independent constant.  The chiral anomaly in Eq.~\eqref{eq:eom_au} in
the presence of $B\neq 0$ reads,
\begin{equation}
  \rn(u^{5/2} \cB^{1/2} a_0')  + 4\alpha B f_{0z} = 0\,.
\end{equation}
Because $f_{0z}=\rn a_z$ (dropping $\rz$), the above two equations are
summarized into
\begin{equation}
  u^{5/2}\cB^{1/2} a_0' + 4\alpha B a_z = c\,,
  \label{eq:constraint}
\end{equation}
where $c$ is a $t$ and $u$ independent constant.  We note that, unlike
the $B=0$ case, $a_0$ takes a nonvanishing value.  Physically
speaking, $u^{5/2} \cB^{1/2} a_0'$ is proportional to the matter
chirality, whilst $B a_z$ is the magnetic helicity up to an overall
factor.  We can interpret the chiral anomaly as a conservation law of
the matter chirality and the magnetic helicity.  It should be noted
that the magnetic helicity plays an important role in the description
of magneto-hydrodynamical evolutions~\cite{Hirono:2015rla}.  Now it is
clear that $c$ physically means a net chirality charge in the system
and it should be, in principle, fixed by an initial condition.

The longitudinal EoM is
\begin{equation}
  -u^{-1/2} f^{-1} \cB^{1/2} \rn f_{0z}
  +\ru ( u^{5/2} f \cB^{1/2} a_z' )
  + 4\alpha B a_0' = 0\,,
  \label{eq:eom_long}
\end{equation}
and we can eliminate $a_0$ by combining Eqs.~\eqref{eq:constraint} and
\eqref{eq:eom_long}, so that we can find a differential equation for
$a_z$ only.  Then, we convert the equation into the one in frequency
space.  The resultant differential equation reads:
\begin{equation}
  \begin{split}
    &\xi^{-3/2} \frac{\Omega^2}{1-\xi^3} \cB^{1/2} a_z + \rxi
    \bigl[ \xi^{-1/2} (1-\xi^3) \cB^{1/2} \rxi a_z \bigr] \\
  &\qquad\quad - 16\alpha^2 (\cB_0-1) \cB^{-1/2} \xi^{1/2}
  (a_z - 1 - \bar{c}) = 0\,.
  \end{split}
  \label{eq:eom_az_freqsp}
\end{equation}
Here, $\bar{c}$ represents a Fourier transform of the sum of
$c/(4\alpha B)$ and the zero mode of $a_z$, which should be singular
as $\delta(\Omega)$ since $c$ as well as $B$ is time independent.
It should be noted that $a_z-1$ is of order $\Omega$ in our choice of
the normalization and so this combination is free from the zero mode.
Thus, in the SSM at finite $B$,  the longitudinal electric
conductivity diverges.  This conclusion is consistent with
Ref.~\cite{Lifschytz:2009si}.

We can give an intuitive physical interpretation to
$\bar{c}\propto \delta(\Omega)$.  In the strict limit of $\Omega=0$,
we are looking at the long time behavior of physical observables, and
then the electric conductivity must diverge in this model.  The reason
is quite simple:  quarks are massless in the SSM, and there is no
other process to destroy chirality.  Thus, the matter chirality
results from the chiral anomaly and eventually blows up under the long
time limit.  In other words, due to the chirality production, the
electric carriers increase with increasing time.   Then, the CME
current grows up linearly as a function of time, and the electric
conductivity corresponding to the linear time dependence is divergent
by definition.

This argument implies that a nonzero $\Omega$ piece in $a_z$ could
still be well-defined.  Even though the strict zero mode is singular, let us keep our
normalization of $a_z\to 1$ at $\Omega\to 0$ for convenience and
$a_z-1$ in the above expression is a contribution from the nonzero
mode.  It is now quite interesting that our calculations can evade a
pathological singularity as long as $\Omega\neq 0$ (including
$\Omega\to 0^+$ for strictly static $B$), for which we can drop
$\bar{c}$.  We can also give an intuitive interpretation to dropping
$\bar{c}$ in physical terms.  We can drop the zero mode if $c$ in
Eq.~\eqref{eq:constraint} happens to cancel $a_z$, which occurs when
the zero mode of the matter chirality is forced to be zero.  In fact,
this matter chirality directly couples to the chiral anomaly, and it
should be reasonable to \textit{define} a finite Ohmic part of the
electric conductivity by imposing an extra condition to neutralize the
matter chirality.  This is our working definition of the Ohmic
electric conductivity.

\begin{figure}
  \centering
  \includegraphics[width=0.95\columnwidth]{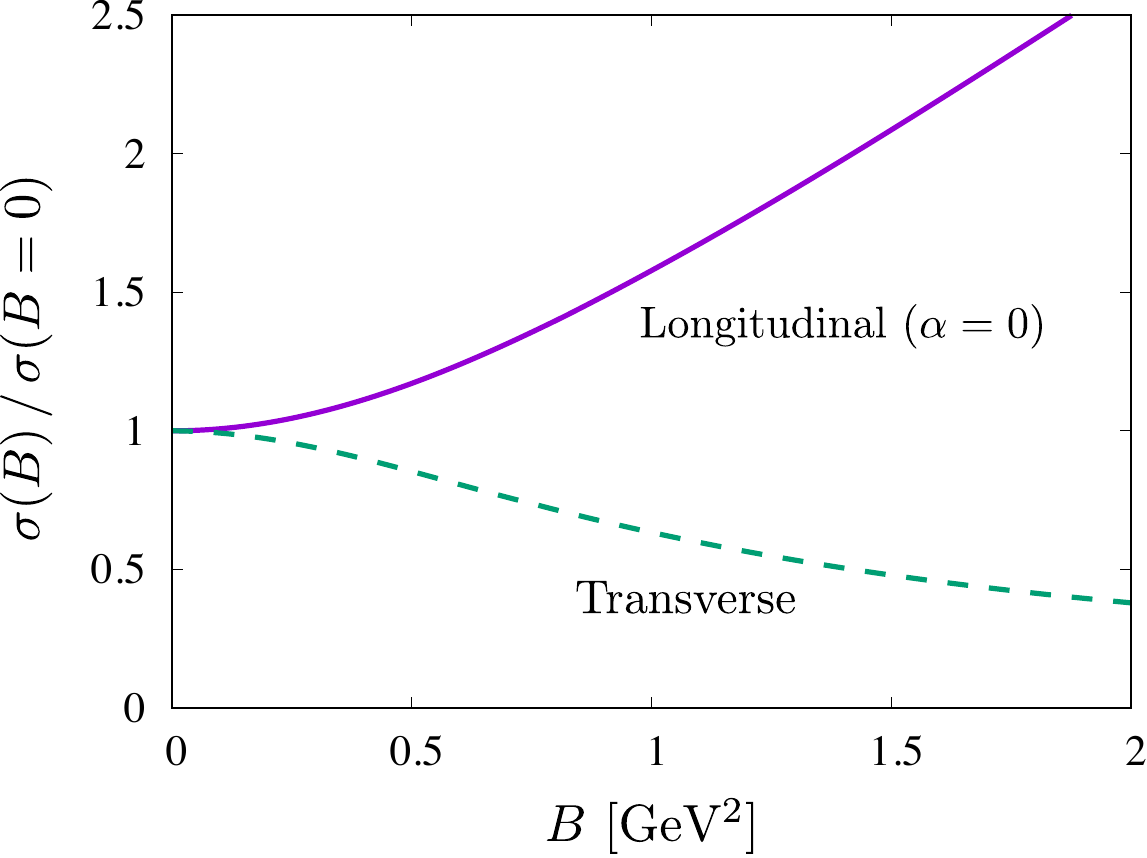}
  \caption{Magnetic dependence of the transverse electric conductivity
    $\sigma_\perp$ and the longitudinal electric conductivity
    $\sigma_\parallel$ at $\alpha=0$.  The physical scale is set at
    $T=0.2\,\text{GeV}$.}
  \label{fig:mag}
\end{figure}

Let  us try to evaluate the electric current under the condition of
$\bar{c}=0$.  It is difficult to find an analytical expression of
$a_z(\xi)$ in general, but the calculation is quite simple in the
$\alpha=0$ case, that is the case with the full suppression of the
chiral anomaly.  In this special limit of $\alpha=0$, first, we can
easily solve the differential equation as
$a_z^{(0)}(\xi)=a_z(\xi;\alpha=0)=(1-\xi)^{-\frac{i\Omega}{3}}
g(\xi;\alpha=0)$ with
\begin{align}
  a_z^{(0)}(\xi)
  &= C\int_0^\xi d\xi\,\frac{\xi^{1/2} \cB^{-1/2}}{1-\xi^3} + D \notag\\
  &\simeq 1 + \frac{2i\Omega}{3}\cB_0^{1/2} \xi^{3/2} + \cdots
  \label{eq:a_z0}
\end{align}
with $C=i\Omega\cB_0^{1/2}$ and $D=1$.  We see that the difference
from $a_x(\xi)$ in Eq.~\eqref{eq:ax_integ} is only the power of $\cB$,
and it is almost obvious that the magnetic dependence is
\begin{equation}
  \sigma_\parallel (\alpha=0) = \sigma(B=0) \sqrt{1+B^2 u_T^{-3}}\,.
  \label{eq:s_mag_l}
\end{equation}
Therefore, in this special case with $\alpha=0$ the longitudinal
conductivity is enhanced by the effect of increasing $B$.  Now, to
see quantitative behavior in the physical units, we convert
$B^2 u_T^{-3}$ into a GeV quantity using
\begin{equation}
  B^2 u_T^{-3} = 9 \biggl(\frac{4\pi}{3}\biggr)^{-4}
  \frac{\Mkk^2 \tilde{B}^2}{\lambda^2 T^6}\,,
\end{equation}
where $\tilde{B}$ is the physical magnetic field.  In
Fig.~\ref{fig:mag} we plot $\sigma(B)/\sigma(B=0)$ (where the tilde is
omitted) for the transverse conductivity in Eq.~\eqref{eq:s_mag_t} and
the longitudinal conductivity at $\alpha=0$ in
Eq.~\eqref{eq:s_mag_l}.  From this it is evident that the modification
is significant for $B$ at the order of $\text{GeV}^2$.
Two curves remarkably match with the numerically estimated results in
Ref.~\cite{Astrakhantsev:2019zkr}, see Fig.~2 of the cited paper.

\begin{figure}
  \centering
  \includegraphics[width=0.9\columnwidth]{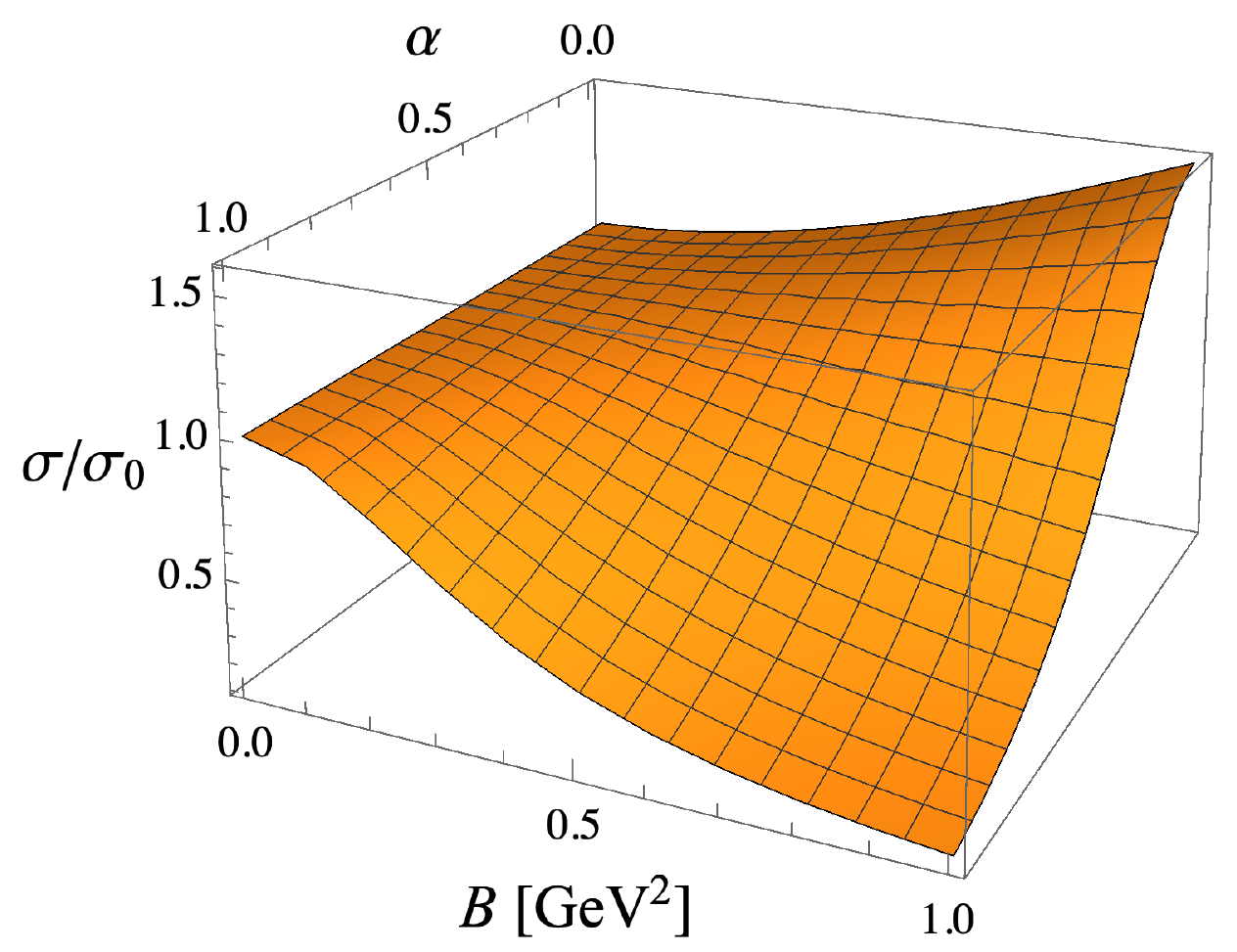}
  \caption{Longitudinal electric conductivity $\sigma_\parallel$
    normalized by its value $\sigma_0$ at $\alpha=B=0$.  The physical
    value of $\alpha$ is $3/4$, for which $\sigma_\parallel$ decreases
    with increasing $B$.}
  \label{fig:conduct}
\end{figure}

Next, we can consider the full $\alpha$ dependence numerically.  For
actual procedures it is convenient to introduce a function,
$\eta(\xi)=(1-\xi)\partial_\xi a_z(\xi)$, and then the infalling
boundary condition can be expressed as
$\eta(\xi\sim 1) = \frac{i\Omega}{3} a_z(\xi\sim 1)$.  The set of two
differential equations can be integrated with an initial condition,
$a_z(\xi\sim 0)=1$, and $\eta(\xi\sim 0)$ should be fixed to satisfy
the infalling boundary condition.  We performed the numerical
calculation by means of the Shooting Method for various $\alpha$ and
$B$, and the results are summarized in Fig.~\ref{fig:conduct}.

For $\alpha=0$ our numerical results in Fig.~\ref{fig:mag} correctly
reproduce the increasing behavior as in Fig.~\ref{fig:conduct}.  Also,
there is no $\alpha$ dependence at all for $B=0$ since the
Chern-Simons action has no contribution, which is confirmed in
Fig.~\ref{fig:conduct}.  It is intriguing to observe that the
qualitative tendency of the $B$ dependence is changed as $\alpha$
increases.  Indeed, for $\alpha=3/4$ (i.e., the physical value),
$\sigma_\parallel(B)$ decreases with increasing $B$, and this is our
central finding.

Actually, for $\alpha=0$, we can give a simple account for the
increasing behavior of $\sigma_\parallel$.  In the limit of strong $B$
the LLL approximation should be justified, and the fermion dynamics is
reduced to (1+1) dimensions along the longitudinal direction.  Then,
massless fermions cannot scatter in (1+1) dimensions (see discussions
in Ref.~\cite{Fukushima:2015wck}) and the transport coefficients are
inevitably divergent~\cite{Hattori:2016cnt}, see more
specifically Fig.~3 in Ref.~\cite{Fukushima:2017lvb}.  This phase space
argument has nothing to do with the chiral anomaly, so that it is
applied to the $\alpha=0$ case.  At the algebraic level we can
understand $\sigma_\parallel\to \infty$ at strong $B$ from
Eq.~\eqref{eq:eom_az_freqsp}.  For $\alpha=0$ and small $\Omega$, the
differential equation to be solved corresponding to
Eq.~\eqref{eq:reduced_eq} is
\begin{equation}
  \rxi \Bigl[ \xi (1-\xi^3) \rxi a_z \Bigr] = 0
\end{equation}
after we replace $\cB\to B^2 u_T^{-3} \xi^3$.  The integration near
$\xi\sim 0$ is singular, which makes $\sigma_\parallel$ divergent.

The situation is drastically changed by the third term
$\propto \alpha^2$ in Eq.~\eqref{eq:eom_az_freqsp}.  In the large $B$
limit, again, the differential equation simplifies and the general
solution can be expressed in terms of the hypergeometric functions.
To meet the boundary condition near $\xi\sim 1$, the conductivity
should come along with a normalization factor that is suppressed by
$\alpha$.  The third term in Eq.~\eqref{eq:eom_az_freqsp} was
originally $4\alpha B a_0'$ and this is proportional to the matter
chirality [i,e., the first term in Eq.~\eqref{eq:constraint}].  It is
therefore the matter chirality that allows for fermion scatterings
even at strong $B$.  We have subtracted the zero mode (and divergent)
contribution from the chiral anomaly, and yet, the nonzero mode (that
is, $a_z-1$ is of order $\Omega$) still plays a role.  This is a
sensible scenario;  the anomaly can generate the chirality, which in
turn means that the chirality can decay via the anomaly.  This is
extremely interesting.  We identified the Ohmic electric conductivity,
but its properties reflect interactions induced by the chiral
anomaly.  A very favorable feature is that the anomaly dependence in
the Ohmic part is \textit{opposite} to the negative magnetoresistance
expected in the zero mode.

\section{Summary}
\label{sec:summary}

We calculated the magnetic field dependence of the electric
conductivity in deconfined QCD matter using a holographic QCD model,
namely, the Sakai-Sugimoto Model.  For simplicity we considered only
the high temperature environment at $T>\Tc$ and solved the equations
of motion in the presence of external magnetic field $B$.

We first checked how far the model can work quantitatively by
estimating the electric conductivity $\sigma$ at $B=0$.  Because of a
mass scale, the Kaluza-Klein mass $\Mkk$, the $T$ dependence is found
to be $\sigma\propto T^2/\Mkk$, but as long as $T\gtrsim \Tc$, we have
verified that our estimates agree well with the lattice-QCD values.

We then proceed to the finite $B$ case, and we found that the
transverse conductivity, $\sigma_\perp$, is suppressed by larger $B$,
which is understandable from the Landau quantization picture.  In
contrast, the longitudinal conductivity, $\sigma_\parallel$, is an
increasing function of $B$ if we drop the Chern-Simons action with
$\alpha=0$.  This is also intuitively understandable from the phase
space argument in the lowest Landau level approximation.  Massless
fermions cannot scatter in effectively reduced (1+1) dimensions, and
transport coefficients generally diverge.  However, our numerical
results for $\alpha\neq 0$ show a turnover; that is,
$\alpha_\parallel$ decreases with increasing $\alpha$ and $B$.  We
gave a plain explanation on this numerical observation.  That is, the
zero mode contribution from the chiral anomaly yields the negative
magnetoresistance (and it is divergent for massless fermions unless a
relaxation time is introduced), and the nonzero mode contribution from
the chiral anomaly can significantly affect the fermion interactions
and even the Ohmic part of the electric conductivity.  Fortunately,
however, the $B$ dependence that we discovered in the Ohmic part is
opposite to the negative magnetoresistance, and it would not impede a
common interpretation of the negative magnetoresistance as a signature
for the chiral magnetic effect.

There are several interesting directions for future investigations.
In the present work we did not include a finite density effect, but
the introduction of the chemical potential is feasible enough.
Another improvement is to generalize the formulation to lower
temperatures in the confined phase.  In this case one needs to solve
the equation of motion for $x_4$, and the calculations become
technically involved, but still possible.

For more quantitatively serious discussions, we should compare the
zero mode and the nonzero mode contributions, and for this, an
extension to massive fermions is needed.  Instead of it, one might
think of introducing a parameter corresponding to a relaxation time in
the equations of motion by hand, but the relaxation time may have
nontrivial dependence on $B$, and such a handwaving treatment would
loose the predictive power.  In fact, the holographic model we
employed here was the top-down one and we believe that our results are
robust in some particular limit of QCD\@.  In the bottom-up approach,
on the other hand, some $B$ dependence may be hidden in model
parameters and assumed geometries, and the predictive power would be
limited.

During completion of the manuscript, we have learned a very
interesting result in Ref.~\cite{Sogabe:2021wqk};  they found a
\textit{positive} magnetoresistance from hydrodynamic fluctuations.
The claim seems to be consistent with what we found in
Fig.~\ref{fig:conduct} at $\alpha=3/4$.  It is an interesting question
whether their mechanism is totally distinct or has some connection to
ours.

\acknowledgments
The authors are grateful to
Irina~Aref'eva,
Karl~Landsteiner,
Shu~Lin,
Kostas~Rigatos
for comments.
AO would also like to thank Shigeki~Sugimoto for useful discussions on 
the D branes in his model. 
This work was supported by Japan Society for the Promotion of Science
(JSPS) KAKENHI Grant Nos.\ 18H01211, 19K21874 (KF),
and No.\ 20J21577 (AO).

\bibliography{adsconduct}
\bibliographystyle{apsrev4-1}

\end{document}